\documentclass[reprint,superscriptaddress,aps,prx]{revtex4-2}

\usepackage[usenames,dvipsnames]{xcolor}
\usepackage[utf8]{inputenc}
\usepackage{bm}
\usepackage{amsfonts}
\usepackage{graphicx,psfrag}

\usepackage{booktabs}
\usepackage{amsmath}
\usepackage{calc}
\usepackage{makecell}
\usepackage{ulem}

\usepackage{nicefrac}

\usepackage{enumitem}
\usepackage{float}
\usepackage{natbib}
\usepackage{tikz}
\usepackage{soul,xcolor}
\usepackage[colorlinks=true,linkcolor=Blue,citecolor=Blue,urlcolor=Blue]{hyperref}

\usepackage{xfp}

\usepackage[caption=false]{subfig}
\usepackage[capitalise]{cleveref}

\def\be{\begin{equation}}
\def\ee{\end{equation}}
\def\bea{\begin{eqnarray}}
\def\eea{\end{eqnarray}}

\DeclareUnicodeCharacter{03BA}{\ensuremath{\kappa}}

\begin{document}

\title{Ignition criteria for trigger waves in cell signaling}
\author{Brandon H. Schlomann}
\email{bschloma@berkeley.edu}     
\affiliation{Department of Molecular and Cell Biology, University of California, Berkeley, CA, USA}
\author{William S. DeWitt}
\affiliation{Department of Genome Sciences, University of Washington, Seattle, WA, USA}
\author{Yuanzhao Zhang}
\affiliation{Santa Fe Institute, Santa Fe, NM, USA}
\author{Kasturi Shah}
\affiliation{Department of Applied Mathematics and Theoretical Physics, University of Cambridge, UK}

\begin{abstract} 
    To rapidly coordinate collective behaviors in tissues, cells communicate with one another through traveling fronts of signaling activity called trigger waves. The stimulus threshold for wave propagation is a key biological parameter that determines when the system produces a global response to a local stimulus. However, it is unclear how this stimulus threshold is determined by the properties of the underlying signaling network. To address this gap, we studied a model of trigger waves with a realistic Hill-type auto-activation function. We obtained an analytic expression for the wave ignition threshold in 1D and numerical solutions in 2D and 3D. In the limit of high sensitivity, we found that the trigger wave threshold depends on both the effective dissociation constant, $K_D$, and the effective Hill coefficient, $n$, of the positive feedback circuit, with the dominant contribution scaling as $K_D^{n/(n-1)}$ in all dimensions. This result leads to a simple method for predicting the trigger wave ignition threshold from bulk activation data and is potentially of use for developing synthetic trigger wave circuits with desired sensitivities.
\end{abstract}

\maketitle

\begin{figure*}
\centering
\includegraphics[width=6in]{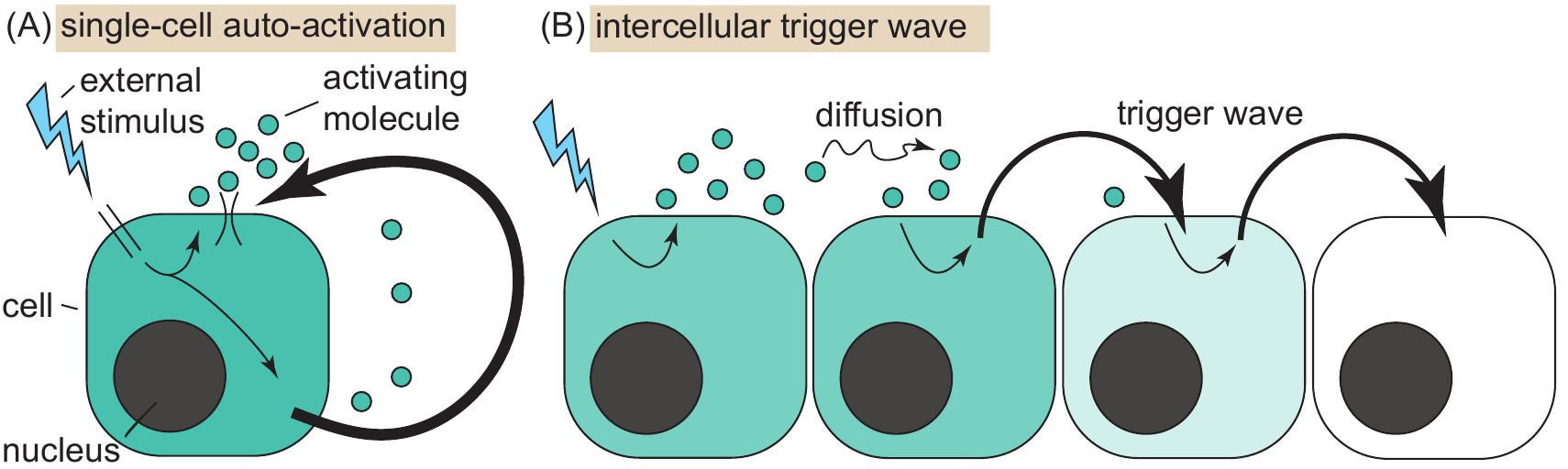}
\caption{\textbf{Trigger waves in cell signaling arise from positive feedback and diffusion of secreted molecules.} (A) Schematic of the positive feedback loop within single cells, which can be activated by an external signal. (B) Schematic of how a localized signal can excite a trigger wave that propagates across a tissue.}
\label{fig:schematic}
\end{figure*}

\section{Introduction}
Populations of cells signal to one another to coordinate collective behaviors. The secretion and diffusion of messenger molecules is a common mechanism of communication between cells \cite{alberts_molecular_2022}, but is limited by the fundamental scaling of diffusive transport through the ambient fluid, $x \sim \sqrt{Dt}$, where $x$ is space, $t$ time, and $D$ the diffusion coefficient \cite{einstein_uber_1905}. For proteins in water, $D \approx 100\mu$m$^2$/s \cite{phillips_physical_2012}, so diffusion over a cell length of approximately $10 \mu$m occurs within a second, but diffusion across a centimeter-sized tissue takes days \cite{gelens_spatial_2014}. Thus, coordinating the activity of cells in tissues on faster timescales requires additional mechanisms. 

One such mechanism is the ignition of signaling trigger waves \cite{gelens_spatial_2014}. Here, traveling waves in the concentration of a signaling molecule (which can be proteins, ions, or other small molecules) are generated and travel across the system at constant speed, $x\sim c t$, enabling cell-cell communication on millimeter and larger length scales much faster than is possible by diffusion alone. Examples of trigger waves in cell signaling include waves of mitosis in developing embryos \cite{hara_cinematographic_1971, chang_mitotic_2013, vergassola_mitotic_2018}, calcium waves in response to wounds \cite{cornell-bell_glutamate_1990}, waves in ERK signaling during tissue regeneration \cite{fan_ultrafast_2023}, and waves of coordinated cell death \cite{co_emergence_2024}. In addition, trigger waves have been predicted to occur in innate immune signaling pathways such as the TNF pathway \cite{yde_analyzing_2011, yde_modeling_2011}. 

Trigger waves arise from a combination of diffusion and positive feedback loops in signaling pathways that exhibit bistability (\cref{fig:schematic}A). Cells can exist in one of two stable states, marked by a low and a high concentration of an activating molecule. When one cell gets activated by a local stimulus, it secretes a molecule that diffuses toward and can activate the neighboring cell. If the stimulus is sufficiently strong, each cell will activate its neighbor in a chain reaction, producing a traveling front (\cref{fig:schematic}B). When additional elements are added to the signaling pathway, such as negative feedback, more complex behaviors can arise, including transient pulses and oscillations \cite{gelens_spatial_2014}. A formula for the wave speed, deducible up to a prefactor from dimensional analysis, has been known for over a century \cite{luther_ii_1906} and provides a useful estimate for understanding the role of trigger waves in real systems \cite{gelens_spatial_2014}: $c = 2\sqrt{D/\tau_p}$, where $D$ is the diffusion coefficient and $\tau_p$ is the timescale associated with the positive feedback. 

Unlike the wave speed, how the stimulus threshold for wave ignition depends on the underlying biochemical signaling is much less well understood. This threshold is an important biological parameter as it determines when the system produces a macroscopic, global response to a microscopic, local stimulus. Especially in contexts such as coordinated cell death \cite{co_emergence_2024} or immune activation \cite{yde_analyzing_2011}, global activation of the tissue via a trigger wave has dramatic physiological consequences and so must be induced with care. At the same time, individual cells must retain some level of sensitivity in order to respond to local stimuli. How cells effectively tune their trigger wave activation threshold to achieve this balance is unclear. In addition, with continued advances in synthetic biology, there is increased interest in designing multicellular circuits that can be programmed to perform specific functions in spatially structured environments such as tumors \cite{trentesaux_harnessing_2023}. A better understanding of how to tune both cell- and tissue-scale sensitivities in bistable signaling networks is crucial for such efforts.   

In this work, we study a minimal but biochemically relevant model of bistable trigger waves. Our work builds on previous works on analytical results for the trigger wave ignition criterion \cite{mornev_conditions_1981, mckean_threshold_1985, flores_stable_1989, neu_initiation_1997, idris_analytical_2008, bezekci_fast-slow_2017, bezekci_strength-duration_2020}. These previous studies focused on either general features of the activation threshold \cite{mornev_conditions_1981} or on models with analytically tractable auto-activation functions, especially of the cubic form \cite{mckean_threshold_1985, flores_stable_1989, neu_initiation_1997, idris_analytical_2008, bezekci_fast-slow_2017, bezekci_strength-duration_2020}. Cubic functions capture the core behavior of trigger wave propagation, especially in the context of action potentials. However, it is difficult to map their parameters onto biochemical signaling networks, which are generally characterized by Hill-type activation functions \cite{zhao_bistable_2023, tran_precision_2018}. Therefore, here we focus on a more realistic model with a Hill function governing auto-activation, at the expense of restricting our analysis to time-independent stimuli.

This paper is organized as follows. First, we introduce our partial differential equation (PDE) model and establish some of its basic properties. Second, we study the threshold behavior of the model without diffusion, reducing the PDE to an ordinary differential equation (ODE), which represents the average behavior of a well-mixed population of cells. Third, we study the full model and compute the critical stimulus threshold required to activate a spatially-extended population of cells with a localized stimulus, analytically in the one-dimensional version of our model and numerically in the two- and three-dimensional versions. We find that in all spatial dimensions the activation threshold scales the same way with the dissociation constant of the auto-activation circuit. This result leads to a simple method for predicting the trigger wave ignition threshold from straightforward measurements of bulk activation in well-mixed environments, which may be useful for understanding when to expect trigger waves to occur in natural and synthetic signaling systems. 

\section{Results}

\begin{figure*}[htb]
\centering
\includegraphics[width=7in]{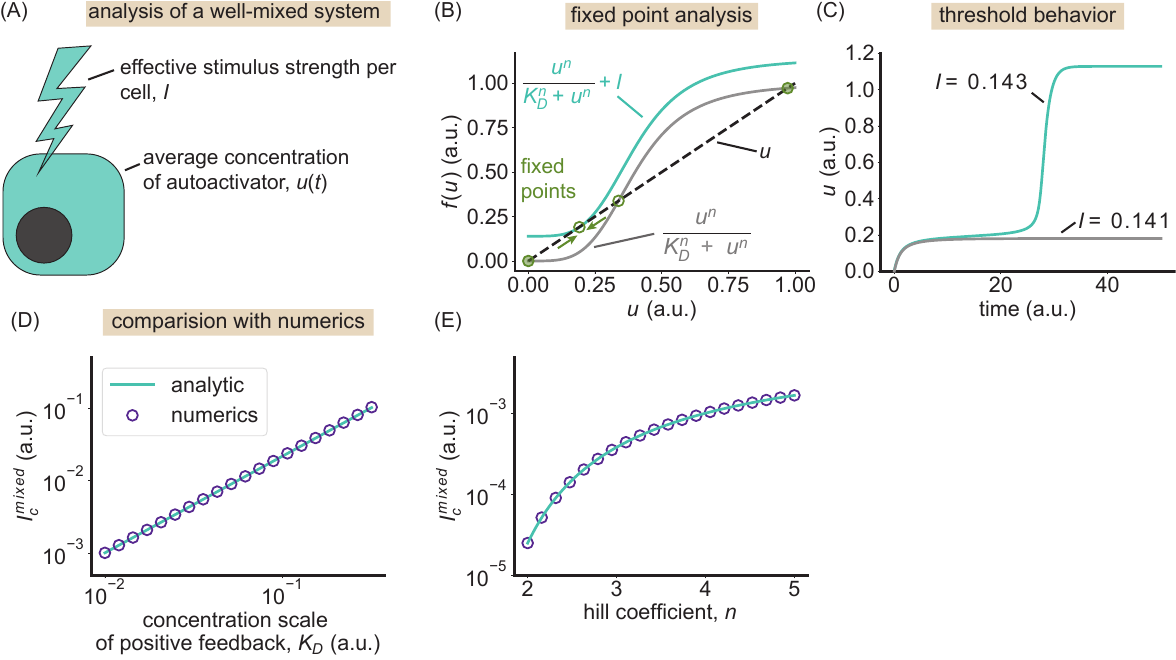}
\caption{\textbf{Analytic results for well-mixed activation agrees well with simulation.} (A) Schematic of the well-mixed problem. Assuming uniform stimulation, we can ignore diffusion in \cref{eq:pde} and analyze the ODE in \cref{eq:ode} for the average response driven by an effective stimulus strength per cell, written here schematically as $I$, see \cref{eq:ode} for details. We seek the critical value of $I$ that activates the system into its high stable state. (B) The key idea of our calculation is to study the fixed point structure of the activation function, $f(u)$. Beyond a critical value of the stimulus strength, the system will lose its low stable fixed point and $u(t)$ will always go to the activated high state. The fixed points of $f(u)$ are shown in green (stable points filled). (C) Example response for $I$ below (gray line) and above (cyan line) the critical threshold. (D)-(E) Comparison between the analytical and numerical results for the critical threshold, $I^\text{mixed}_c$, as a function of model parameters. In (D), $n=4$ and in (E), $K_D=0.01$.}
\label{fig:hill_analytics_mixed}
\end{figure*}

\subsection{Model definition and basic properties}
\label{sec:model_def}

We begin by considering a reaction-diffusion system describing the concentration of an activating molecule, $u(\vec{x},t)$, on a $d$-dimensional lattice containing many cells (\cref{fig:hill_analytics_mixed}A). We model a generic auto-activating signaling network by combining an effective Hill function describing the positive feedback loop with a first-order decay process \cite{co_emergence_2024, sokolik_transcription_2015, zhao_bistable_2023}, coupled in space by diffusion. We consider subjecting one cell, located at the origin, to a localized stimulus with rate $2I$ (units of amount of $u$ per time; the factor of 2 is conventional \cite{idris_analytical_2008}). Putting these ingredients together and taking the continuum limit, we obtain,
\be
\partial_tu = D\nabla^2u + k\frac{u^n}{K^n_D + u^n} - \gamma u + 2\,(l_{cell})^d \,I\, \delta(\vec{x}).
\label{eq:pde}
\ee
Here, $D$ is the diffusion coefficient, $k$ is the maximum production rate, $n$ is the Hill coefficient that sets the sharpness of the auto-activation circuit, $K_D$ is the effective dissociation constant that sets the sensitivity of the auto-activation circuit (it is concentration of $u$ at which the positive feedback becomes relevant), and $\gamma$ is the decay rate. The effective Hill function parameters, $n$ and $K_D$, describe the response of the full signaling network---which may include many intermediate steps---to auto-activation and can in principle be determined empirically by exogenously varying the concentration of the activator and measuring the response. The localized stimulus at the origin is described by the Dirac delta function, $\delta(x)$, with the prefactor $(l_{cell})^d$ representing the lattice voxel (see \cref{sec:pde} for details; keeping track of lattice size is useful for our comparison to the non-spatial model in \cref{sec:mfa}). We restrict our analysis to parameter regimes where the model locally exhibits bistability, with low-$u$ and high-$u$ stable states. 

Our goal is to calculate the critical threshold, $I^\text{wave}_c$, that will trigger a traveling wave solution in \cref{eq:pde} and to understand how different biochemical parameters influence the trigger wave threshold. To simplify the analysis, we non-dimensionalize \cref{eq:pde} via $u \to u / (k/\gamma)$, $t \to \gamma t$, $\vec{x} \to \vec{x} / \sqrt{D/\gamma}$. The resulting dimensionless governing equation is
\be
\partial_t u = \nabla^2 u + \frac{u^n}{K^n_D + u^n} - u + 2\,(l_{cell})^d\,I\,\delta(x),
\label{eq:pde_nodim}
\ee
where $K_D$ and $(l_{cell})^dI$ have been rescaled appropriately. This model only has three parameters: the dimensionless dissociation constant, $K_D$, the Hill coefficient, $n$, and the dimensionless stimulus strength, $(l_{cell})^dI$. We also define the activation function
\be
f(u) := \frac{u^n}{K^n_D + u^n} - u  
\ee
and its three fixed points, $u_1 < u_2 < u_3$. 

As we will see below, the structure of $f(u)$ and the values of its fixed points---especially $u_2$---determine much of the behavior of \cref{eq:pde}. Therefore, we calculate them here by setting $f(u)=0$ and solving for $u$. The low activity state is trivially $u_1 = 0$. To make analytical progress on $u_2$ we work in the limit $K_D\ll 1$. This approximation corresponds to a limit of high sensitivity, where the concentration scale of $u$ that induces positive feedback is small compared to the equilibrium value of $u$ in the high state. This approximation is reasonable because $K_D$ generally must be small ($< 1$) for the system to be bistable. The intermediate fixed point, $u_2$, satisfies
\be
u^{n-1}_2 - u^n_2 = K_D^n.
\ee
Since $u_2$ is set by $K_D$, which we take to be small, we conjecture that $u_2$ is also small and will check this conjecture with \textit{a posteriori} reasoning. Assuming that $u_2$ is small, we can neglect the $u^n_2$ term and find that
\be
u_2 \approx K^{\frac{n}{n-1}}_D
\label{eq:u2}
\ee
plus higher-order terms, which are indeed small for $n > 1$. By similar reasoning, the high fixed point, $u_3$, is given by $u_3 \approx 1 - K^n_D$ for $K_D \ll 1$. With these fixed points computed, we can begin to understand the behavior of \cref{eq:pde}.

\subsection{Well-mixed analysis}
\label{sec:mfa}

To gain intuition about how the Hill function parameters, $K_D$ and $n$, control the nonlinear response of the system, we begin by studying the mean-field behavior of \cref{eq:pde}, describing the average response of a well-mixed population of cells (\cref{fig:hill_analytics_mixed}A). To do this, we consider the ODE counterpart to \cref{eq:pde},
\be
\dot{u} = k\frac{u^n}{K^n_D + u^n} - \gamma u + 2I.
\label{eq:ode}
\ee
Here, all cells are subject to the same driving stimulus of strength $2I$ (units of amount of $u$ per time; defining $I$ as we did for \cref{eq:pde}, with the factor of $l_{cell}$ scaled out, allows us to directly compare the spatially-dependent and non-spatial models). For this system, the critical stimulus threshold for activation, $I_c^{\text{mixed}}$, corresponds to the value of $I$ that eliminates the low-$u$ stable fixed point in \cref{eq:ode} (\cref{fig:hill_analytics_mixed}B-C; \cref{sec:single-cell}). In the low $K_D$ limit, we use a result for the discriminant of a special set of $n^{th}$ order polynomials \cite{mora_discriminant_2012} to compute this threshold to be
\be
I_c^\text{mixed} =\frac{1}{2}\left(1-\frac{1}{n}\right)\left(\frac{1}{n}\right)^{\frac{1}{n-1}} K_D^{\frac{n}{n-1}},
\label{eq:single-cell}
\ee
or, in dimensionful units, 
\be
I_c^\text{mixed} =\frac{1}{2}\left(1-\frac{1}{n}\right)\left(\frac{1}{n}\right)^{\frac{1}{n-1}} \left(\frac{K_D}{k/\gamma}\right)^{\frac{n}{n-1}}k
\label{eq:single-cell_dims}
\ee
(see \cref{sec:single-cell} for details). This result agrees well with simulations over a wide range of parameters (\cref{fig:hill_analytics_mixed}D-E). Intuitively, this result tells us that decreasing the sensitivity (increasing $K_D$) and increasing the sharpness (increasing $n$) of the auto-activation circuit make it harder to activate the system. 

The key takeaway from our result is the scaling of the stimulus threshold with $K_D^{n/(n-1)}$, which describes how the dissociation constant, $K_D$, and the Hill coefficient, $n$, of the auto-activation pathway combine to determine the nonlinear response of the system. It is interesting to note that the stimulus threshold inherits its dependence on $K_D$ from the intermediate fixed point of the activation function  (\cref{eq:u2}). We next seek to understand if this mean-field behavior also controls the ignition of trigger waves in spatially-extended systems.

\begin{figure*}[htb]
\centering
\includegraphics[width=7in]{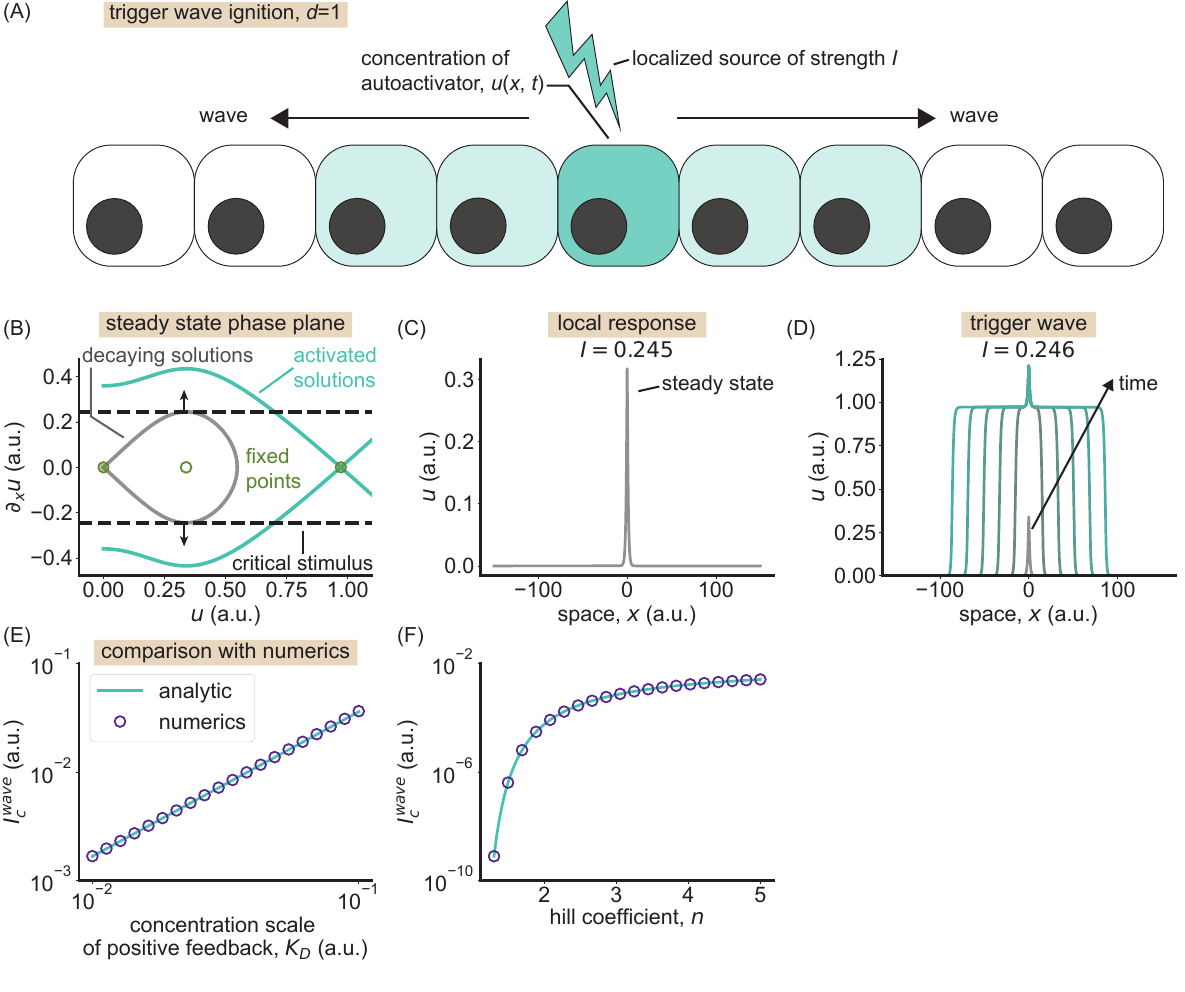}
\caption{\textbf{Analytic result for trigger wave ignition criteria agrees well with simulation.} (A) Schematic of the spatially-extended problem. We consider a 1D grid of cells that produce and consume an auto-activating molecule, $u(x,t)$. One cell at the origin is subjected to a constant stimulus of strength $I$. We seek the critical value of $I$ that ignites a wave of $u$ activity. (B) The idea of the calculation is to study the existence of steady state solutions in the $u-u'$ plane. Solutions that decay in space, with $u(x\to\pm\infty) = u'(x\to\pm\infty) =0$ lie on the solid gray line. Solutions that are uniformly high in the activated state lie on the solid cyan line. The boundary conditions on $u'(x=0^{\pm})$ are shown as horizontal black dashed lines. Beyond a critical value of the stimulus strength, $I$, the boundary condition at the origin can only be satisfied by the teal (activated) solution set, corresponding to the onset of trigger waves. The three fixed points, $u_1 < u_2 < u_3$, of the activation function, $f(u)$, are shown in green (stable points filled). The critical stimulus value corresponds to the point $u'(u_2)$ on the gray (decaying) solution set. (C) Example of a local steady state response for $I$ below the critical threshold. (D) Example of a trigger wave response above the critical threshold. (E)-(F) Comparison between the analytic and numerical results for the critical threshold, $I^{wave}_c$, as a function of model parameters ($l_{cell}$ absorbed into the definition of $I$). For (E) $n=4$ and for (F), $K_D=0.01$.} 
\label{fig:hill_analytics}
\end{figure*}

\begin{figure*}[htb]
\centering
\includegraphics[width=6.5in]{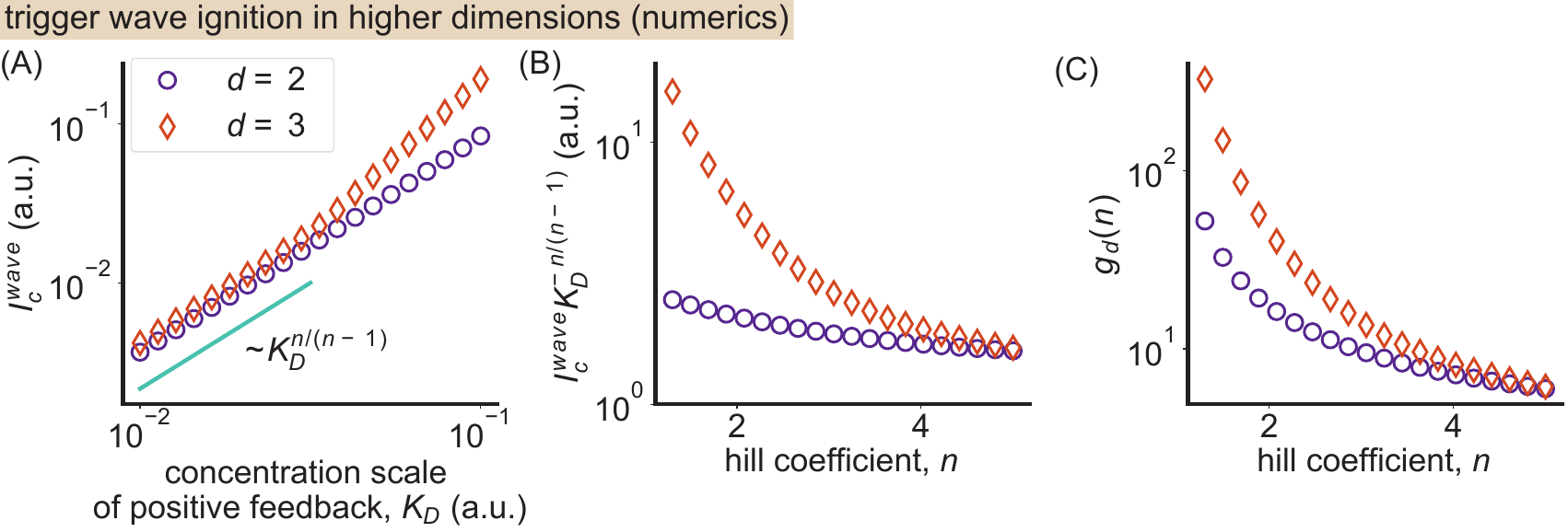}
\caption{\textbf{The scaling of the trigger wave threshold with $K_D$ extends to higher dimensions.} (A) Comparison between the scaling prediction of the critical threshold, $I^{wave}_c$, as a function of $K_D$ for $d=2$ (purple circles) and $d=3$ (orange diamonds; $l_{cell}$ is absorbed into the definition of $I$). The scaling behavior seen in the well-mixed system and in $d=1$ (cyan line) holds in higher dimensions for $K_D \ll 1$. (B) Numerical results for the Hill coefficient-dependent prefactor of $I^{wave}_c$. (C) Numerical results for the Hill coefficient-dependent prefactor of $I^{wave}_c / I^{mixed}_c$, defined as $g_d(n)$ in \cref{eq:gd}. For (A) $n=4$ and for (B)-(C), $K_D=0.01$.}
\label{fig:higher_dims}
\end{figure*}

\subsection{Trigger wave ignition threshold}
To compute the wave ignition threshold (\cref{fig:hill_analytics}A) in $d=1$, we follow the approach of Mornev \cite{mornev_conditions_1981} (recently introduced to a wider audience in \cite{bezekci_strength-duration_2020} and also related to approaches outlined in \cite{arnold_vladimir_ordinary_1973}), which applies to arbitrary activation functions with time-independent stimuli in one dimension. This approach exploits the mapping between stationary solutions of one-dimensional reaction diffusion systems and the dynamics of a single-particle Hamiltonian system \cite{arnold_vladimir_ordinary_1973}. The idea is to identify for what values of $I$ do steady-state solutions of \cref{eq:pde} that decay to zero at infinity, $u(x\to\pm\infty) = u'(x\to\pm\infty) = 0$, cease to exist. 
In \cref{sec:tissue}, we show that such decaying steady-state solutions are constrained by the following relationship between $u(x)$ and its derivative, $u'(x) = \partial_x u$:
\be
u' = \left(-2\int du\,f(u)\right)^{1/2}.
\ee
Drawing this solution set in the $u-u'$ plane, we see that it forms a closed curve that starts and ends at $(0, 0)$ (\cref{fig:hill_analytics}B, gray curve). Traveling along this curve clockwise outlines the shape of a localized steady state profile (\cref{fig:hill_analytics}C). In contrast, solutions with spatially uniform activation contain the point $(u, u') = (u_3, 0)$, with $u_3$ the high fixed point of $f(u)$ and $u'=0$ describing a flat activation profile; this solution set is drawn in  \cref{fig:hill_analytics}B as a cyan curve (see \cref{sec:tissue} for details). 

Since the Delta function stimulus in \cref{eq:pde} fixes the value of $u'$ at $x=0$, we can draw this boundary condition as a pair of horizontal lines in the $u-u'$ plane (one for $\lim_{x\to  0^+}$ and one for $\lim_{x\to 0^-}$; \cref{fig:hill_analytics}B, dashed black lines). As $I$ increases, these horizontal lines move away from the $x$ axis. The critical stimulus strength to ignite a wave, $I_c^\text{wave}$, corresponds to the point at which these horizontal lines lift off of the lower solution branch (indicated by black arrows in \cref{fig:hill_analytics}B; this transition is shown in \cref{fig:hill_analytics}C-D). It can be shown that this point---which corresponds to the apex of the lower branch---occurs at $u=u_2$, the intermediate, unstable fixed point of $f(u)$ (see \cref{sec:tissue}) \cite{mornev_conditions_1981}. Therefore, $I_c^\text{wave}$ can be computed by solving the integral
\be
l_{cell}I_c^\text{wave} = \left(-2\int^{u_2}_0 du\,f(u)\right)^{1/2}.
\ee

We solved this integral to leading order in the limit $K_D\ll 1$, corresponding to a high sensitivity limit (\cref{sec:tissue}). We find the trigger wave threshold to be 
\be
l_{cell}I_c^\text{wave} = \left(\frac{n-1}{n+1}\right)^{1/2}K_D^{\frac{n}{n-1}},
\label{eq:tissue}
\ee
or in dimensional units,
\be
I_c^\text{wave} = \left(\frac{n-1}{n+1}\right)^{1/2}\left(\frac{K_D}{k/\gamma}\right)^{\frac{n}{n-1}}\sqrt{\frac{D}{\gamma \,l^2_{cell}}}\,k.
\label{eq:tissue_dims}
\ee
(Note: our calculation also assumes the continuum limit $\sqrt{D/\gamma} \gg l_{cell}$.) This result agrees well with numerical simulations  across a wide range of parameters (\cref{fig:hill_analytics}E-F); see \cref{sec:numerics} for our numerical scheme.

Our result indicates that in $d=1$, the dependence of the trigger wave ignition threshold on $K_D$ is, to leading order, identical to the case of mean-field activation threshold. We found numerically that the same behavior occurs for $d=2$ and $d=3$ (\cref{fig:higher_dims}A; some noticeable deviation occurs for $d=3$ at larger $K_D$, indicating that the small $K_D$ approximation breaks down sooner here), which suggests a universal form of the wave ignition threshold,
\be
I_c^\text{wave} \sim \left(\frac{K_D}{k/\gamma}\right)^{\frac{n}{n-1}}\left(\frac{D}{\gamma \,l^2_{cell}}\right)^{d/2}\,k
\ee
for $K_D/(k/\gamma) \ll 1$.

An interesting consequence of this scaling behavior is that the ratio between the wave and well-mixed thresholds can be written in a surprisingly simple form
\be
\frac{I^\text{wave}_c}{I^\text{mixed}_c} = g_d(n)\left(\frac{l_D}{l_\text{cell}}\right)^d,
\label{eq:gd}
\ee
where $g_d(n)$ contains the dependence on the Hill coefficient, $l_D = \sqrt{D/\gamma}$ is the diffusive length scale, and $l_\text{cell}$ is the size of a cell. In $d=1$, the function $g_d(n)$ is given by 
\be
g_1(n) = \frac{2\left(\frac{n-1}{n+1}\right)^{1/2}}{\left(1-\frac{1}{n}\right)\left(\frac{1}{n}\right)^{\frac{1}{n-1}}},
\ee
which is a decreasing function of $n$ for which $g_1(2)\approx 4.62$ and $g_1(\infty)=2$. In higher dimensions, we find numerically that there is a larger dynamic range of the Hill coefficient-dependent prefactor of $I^{wave}_c$ (\cref{fig:higher_dims}B), which translates to a larger dynamic range of $g_d(n)$ (\cref{fig:higher_dims}C). We find that $g_d(n) \approx 7$ for $n\approx 4$ in $d=2$ and $d=3$, but takes on larger values at smaller $n$, with $g_2(2) \approx 16$, and $g_3(2) \approx 48$. However, in most of parameter space, the dominant contribution to the trigger-wave-to-bulk activation ratio comes from the ratio of diffusive to cellular length scales. As we discuss below, this result suggests a simple method for predicting the ignition threshold of a trigger wave circuit based on straightforward bulk measurements of cellular activity.

\section{Discussion}
In this paper, we studied a reaction-diffusion model of cell signaling trigger waves governed by a Hill function-based auto-activation circuit. When this model is bistable, its behavior is qualitatively similar to the well-studied Nagumo equation, $\partial_t u = \partial^2_x u + u(u-\theta)(1-u)$ \cite{idris_analytical_2008}. The cubic form of the activation function in the Nagumo equation has enabled a wealth of analytic results, including detailed analysis of time-dependent stimuli \cite{idris_analytical_2008, bezekci_strength-duration_2020}. However, in the context of cell signaling trigger waves, auto-activation circuits are characterized by Hill functions, not cubic functions. The key difference between the two is that Hill functions can support higher degrees of nonlinearity than cubic polynomials \cite{zhao_bistable_2023, tran_precision_2018} and also saturate at high input levels. Our contribution in this paper is to understand the quantitative details of how the wave ignition threshold depends on Hill function parameters---the sensitivity, $K_D$, and the sharpness, $n$---that describe real biochemical signaling networks \cite{zhao_bistable_2023, tran_precision_2018}.

Our results have some practical implications for experiments. The ratio of the wave to well-mixed activation thresholds is a simple function whose value is set primarily by the ratio of diffusive to cellular length scales, $l_D/l_{cell} = \sqrt{D/\gamma}/l_{cell}$. Therefore, if one measures the activation threshold in bulk with a population of cells, for example by flow cytometry \cite{gottschalk_distinct_2016}, one can make a prediction for the trigger wave ignition threshold simply by knowing the diffusion coefficient and half-life of the signaling molecule. This simple estimate could be useful for understanding when we should expect to see trigger waves occur in tissues.

For example, trigger waves of the proinflammatory molecule TNF$\alpha$ have been predicted to occur based on the topology of its signaling pathway \cite{yde_modeling_2011, yde_analyzing_2011}. However, despite exhibiting rich spatiotemporal dynamics \cite{son_spatiotemporal_2022}, trigger waves of TNF$\alpha$ have not yet been observed experimentally (some suggestive evidence in a similar pathway in fruit flies does exist \cite{singh_serpin27a-dependent_2021}). One possibility is that, while the TNF pathway exhibits positive feedback, it may not have parameter values that support traveling wave solutions. However, another possibility is that the stimulus threshold to ignite the wave is higher than stimulus levels that have been tested experimentally. Using typical values for a signaling protein, with $D= 100 \mu$m$^2$/s, $\gamma = 3 \cdot 10^{-3}$ s$^{-1}$ (half-life of $\approx$ 5 minutes), and a cell size of 10 $\mu$m, we have that the ratio of diffusive to cellular length scales is approximately 18. Therefore, as a lower bound, we predict that the trigger wave threshold is at least $18$ times that of the well-mixed activation threshold in $d=1$. For higher dimensions, this lower bound on the activation threshold ratio becomes 324 in $d=2$, and approximately 5800 in $d=3$. Since the stimulus strength, $I$, can be interpreted as the initial rate of auto-activator production immediately after stimulation, achieving these levels of fold-change induction is likely difficult for $d\ge 2$. However, whether or not this is actually the scenario for the TNF pathway is difficult to say, as this pathway contains important complexities that are outside of our model.

We suggest that there are two key  elements lacking in our model that constitute natural next steps for this work and will help make contact with pathways like TNF. First is the extension to time-dependent stimuli. Especially in the context of action potentials in neurons, there is a long history of studying the wave ignition threshold as a function of stimuli that are localized in time as well as space \cite{bezekci_strength-duration_2020}. Such time dependence is equally important to consider for cell signaling responses, which are often dealing with dynamic signals \cite{son_spatiotemporal_2022}. Extending our analysis to time-dependent stimuli will be an important direction for future work. Second is the addition of negative feedback to the model. In the context of action potentials, negative feedback is a key part of the celebrated Fitz-Hugh Nagumo model and generates complex behaviors including oscillations \cite{bezekci_fast-slow_2017, bezekci_strength-duration_2020}. In the context of cell signaling, negative feedback is also widespread, including in the TNF pathway, where it produces oscillations of the transcription factor NF-$\kappa$B in and out of the nucleus \cite{yde_analyzing_2011, yde_modeling_2011, son_spatiotemporal_2022, adelaja_six_2021}. Negative feedback was also found to be important for tuning the bistability of a programmed cell death signaling network that supports trigger waves \cite{co_emergence_2024}. It would be interesting to see if the scaling behavior of our results derived here holds with the addition of negative feedback.

\begin{acknowledgments}
BHS,  WSD, and KS were funded by individual fellowships from the James S. McDonnell Foundation. YZ was supported by an Omidyar Fellowship from the Santa Fe Institute. In addition, we gratefully acknowledge the James S. McDonnell Foundation and the Santa Fe Institute for supporting this work through funding for two working groups held at the Santa Fe Institute in March and October 2023. We thank Isa Stallworthy and Hana Mir for helpful discussions, Yasemin Kirişçioğlu for help tracking down the Mornev reference \cite{mornev_conditions_1981}, and Hilary Skolnik, Bruce Bertram, and Jen Leighton for logistical support at the Santa Fe Institute. 
\end{acknowledgments}

\appendix
\section{Deriving a continuum model of immune trigger waves}
\label{sec:pde}

Consider a 1D lattice of cells of size $l_{cell}$ indexed by position $i$ that produce an autoactivator $u$ in response to a stimulus $2I$ at the origin and in response to itself:
\be
\dot{u}_i =  k\frac{u^n_i}{K^n_D + u^n_i} - \gamma u_i + 2 I \delta_{i,0} + \frac{1}{\tau}\left(u_{i+l_{cell}} + u_{i-l_{cell}} - 2u_i\right)
\ee
with $\tau$ the timescale of molecules of $u$ hopping between lattice sites. 
On taking the continuum limit, we obtain
\be
\partial_tu = D\partial^2_xu + k\frac{u^n}{K^n_D + u^n} - \gamma u + 2\,l_{cell}\, I\, \delta(x)
\ee
where $u(x, t)$ is a continuous field and the diffusion coefficient $D \equiv l_{cell}^2/\tau$. The factor of $l_{cell}$ in the stimulus term comes from turning the Kronecker Delta into the Dirac Delta function in the continuum limit. In $d$ spatial dimensions, the $\delta$ function acquires a prefactor of $(l_{cell})^d$, as is required by dimensional analysis. 

\section{Well-mixed analysis}
\label{sec:single-cell}
For analyzing the behavior of a well-mixed population of cells, we will ignore the diffusive flux and just consider the following ordinary differential equation for the reaction terms:
\be
\dot{u} = \frac{u^n}{K^n_D + u^n} -  u + 2 I,\,\,\,\,u(0) = 0.
\label{eq:ode_appendix}
\ee
As explained in the main text (\cref{sec:model_def}), \eqref{eq:ode_appendix} permits three fixed points: one unstable and two stable, where the stable node of lower value is termed the ``low'' state and of higher value the ``high'' state. We seek the critical value $I=I^{mixed}_c$, that drives $u$ to its high state. Since we always have the initial condition $u(0)=0$, we can simply ask what value of $I$ causes us to lose the low stable fixed point, forcing the system to flow to the high fixed point. As in the main text, we work in the ``high sensitivity'' limit $K_D \ll 1$.
 
Setting $\dot{u}=0$, we get the polynomial
\be
-u^{n+1} + (1+2I)u^n - K^n_D I + 2IK^n_D = 0.
\ee
To identify when this system loses its low stable fixed point, we must compute the value of $I$ that causes this polynomial to go from 3 to 2 unique positive real roots, due to the lowest two roots merging. In the main text \cref{sec:model_def}, we show that the second largest root is approximated by $K^{n/(n-1)}_D$ for $K_D \ll 1$. Since $K^{n/(n-1)}_D$ will also be small when $K_D$ is small, we can simplify the problem by dropping the highest power in $u$. Rearranging a bit, we have
\be\label{eqn:polynomial}
u^n - \frac{K^n_D}{1+2I} u + \frac{2IK^n_D}{1+2I} = 0.
\ee
So now we want to know when this polynomial goes from having 2 unique positive real roots, to 1. 

Applying Descarte's Rule of Signs \cite{descartes_discourse_2006}, \eqref{eqn:polynomial} has either 2 or 0 positive real roots.  Therefore,
we are looking for when the two roots become degenerate. 
A polynomial will have degenerate roots when its discriminant is zero. 
While computing discriminants for general $n$-degree polynomials is intractable, our trinomial form 
\eqref{eqn:polynomial}
has an explicit expression for the discriminant.

Theorem 1 from \cite{mora_discriminant_2012} states that for a general trinomial
\be
x^n + c_1x^k +c_2,
\ee
the discriminant is given by
\begin{alignat}{2}
D_f = &n^n(-1)^{\frac{n(n-1)}{2}}c^{k-1}_2 \nonumber \\
&\left[c^{\frac{n-k}{d}}_2 - (-1)^{\frac{n}{d}}c^{\frac{n}{d}}_1\left(1-\frac{k}{n}\right)^{\frac{n-k}{d}}\left(\frac{k}{n}\right)^{\frac{k}{d}}\right]^d,
\end{alignat}
where $d\equiv \text{gcd}(n, k)$. In our case, $k=1$, so $d=\text{gcd}(n, 1) = 1$ and $c_1 < 0$. The term in brackets determines when $D_f=0$, which simplifies to (with $c_1 \to -c_1$ for notational consistency with \eqref{eqn:polynomial})
\be
c_2 = c^{\frac{n}{n-1}}_1\left(1-\frac{1}{n}\right)\left(\frac{1}{n}\right)^{\frac{1}{n-1}}
\ee
Substituting for $c_1$ and $c_2$ from \eqref{eqn:polynomial},
we get
\be
2I(1+2I)^{\frac{1}{n-1}} =\left(1-\frac{1}{n}\right)\left(\frac{1}{n}\right)^{\frac{1}{n-1}} K^{\frac{n}{n-1}}_D.
\ee
Dropping the $(1+2I)$ term, we arrive at the critical single cell threshold
\be
I^{mixed}_c =\frac{1}{2}\left(1-\frac{1}{n}\right)\left(\frac{1}{n}\right)^{\frac{1}{n-1}} K^{\frac{n}{n-1}}_D.
\ee


Restoring dimensional units, we get the final result
\be
I^{mixed}_c =\frac{1}{2}\left(1-\frac{1}{n}\right)\left(\frac{1}{n}\right)^{\frac{1}{n-1}} \left(\frac{K_D}{k/\gamma}\right)^{\frac{n}{n-1}}k.
\ee
Higher-order contributions to $I^{mixed}_c$ can be calculated; we do not pursue this here, as the leading order result appears quite accurate (see \cref{fig:hill_analytics_mixed}E-G of the main text).

\section{Trigger wave ignition threshold}
\label{sec:tissue}
The non-dimensionalized model in $d=1$ reads
\be
\partial_t u = \partial^2_x u + f(u) + 2I\,\delta(x),
\label{eq:pde_appendix}
\ee
with
\be
f(u) = \frac{u^n}{K^n_D + u^n} - u
\label{eq:f_appendix}
\ee
(for simplicity, we absorb the factor of $l_{cell}$ into our definition of $I$). Our goal is to calculate the critical value of $I$ that ignites a trigger wave. We follow the result of Mornev \cite{mornev_conditions_1981}, mentioned more recently by \cite{bezekci_strength-duration_2020} and also related to approaches outlined in \cite{arnold_vladimir_ordinary_1973}. This result says that, for any bistable $f(u)$ with three fixed points, $u_1 < u_2 < u_3$, the critical value of $I$ is given by
\be\label{eqn:I_integral}
I^{wave}_{c} = \left(-2 \int_{u_1}^{u_2}f(u)\,du\right)^{1/2}.
\ee
We derived the fixed points $u_1$ and $u_2$ in the main text \cref{sec:model_def} ($u_1=0$, $u_2\approx K_D^{n/(n-1)}$) and will compute the integral below. Before proceeding with the calculation, however, we give a brief sketch of the derivation. Mornev's original work is in Russian, which none of the present authors can read. Therefore, we give here our own derivation of \cref{eqn:I_integral}, which we believe is related to the original paper. 

\subsection{Deriving Mornev's result}
The idea of this approach to is to study when steady state solutions of \cref{eq:pde_appendix} that decay to zero cease to exist. In the steady state, we have
\be
u'' = -f(u) - 2I\delta(x).
\ee
We approach the analysis by rewriting this equation as two first-order equations,
\bea
u' &=& v\\
v' &=& -f(u),
\eea
and replacing the Delta function stimulus with a boundary condition on $v(x)$ at $x=0$. This system exhibits a constant of motion, or ``energy'', given by
\be
E = \frac{v^2}{2} + U(u)
\ee
with potential function
\be
U(u) = \int du f(u).
\ee
Here, ``constant of motion'' means a function that is constant along the profile of the solution in space. Therefore, we can construct solutions with particular energy values using the relation
\be
u' = \sqrt{2\left(E - U(u)\right)}.
\ee
We are only interested in solutions with zero flux at $x\to\pm\infty$, so they must contain the point $u' = 0$. Solutions that decay to zero must include the point $u=0$. Since $U(0)=0$, the only value of $E$ that contains both $u=0$ and $u'=0$ is $E=0$. Thus, we conclude that steady state solutions that decay to zero are defined by the relation
\be
u' = \sqrt{-2\int f(u)du}.
\label{eq:decaying_solutions_appendix}
\ee

In the main text \cref{fig:hill_analytics}A, we plot this solution set in the $u'-u$ plane as a solid gray curve. It consists of a closed curve that starts and ends at $(u, u') = (0, 0)$ (a homoclinic orbit). In a similar way, we can find the set that corresponds to solutions that contain the point $(u,u') = (u_3, 0)$, which correspond to solutions with uniform high activity ($u_3$ being the high fixed point of $f(u)$). This solution corresponds to $E = \int_0^{u_3}f(u)du$. We plot this solution set in \cref{fig:hill_analytics}A as a solid cyan curve. Finally, since the Delta function stimulus term enforces a boundary condition on $u'(0)$, it can be drawn as horizontal lines in the $u-u'$ plane (dashed black lines in \cref{fig:hill_analytics}). 

By visual inspection, we see that for a critical value of the stimulus strength, $I$, the horizontal lines representing the boundary condition will lift off of the gray solution set corresponding to decaying solutions, leaving only the uniform high-state solution in cyan being a possibility. This transition corresponds to the onset of trigger waves. The point can be computed since it is the apex of the decaying solution curve, so occurs at the maximum of the function \cref{eq:decaying_solutions_appendix}. Differentiating \cref{eq:decaying_solutions_appendix} once, we recover the fact that $u'' = -f(u)$, and so observe that the apex of the curve occurs at $u=u_2$. Evaluating $u'(u_2)$, we arrive at the critical gradient at the origin,
\be
(u')_c = \sqrt{-2\int_{u_1}^{u_2}f(u)du}.
\ee
The last piece is to note that having a stimulus strength of $2I$ creates a gradient at $x=0$ of magnitude $I$ on either side of the origin (hence the factor of 2 in the definition of $I$). To see this, consider integrating the steady state of \cref{eq:pde_appendix} across a small region $(-\Delta, \Delta)$ around the origin and taking the limit $\Delta\to 0$. The reaction term $f(u)$ is continuous across the origin, so does not contribute. We are left with a jump condition on the gradient:
\be
u'(0^+) - u'(0^-) = -2\,I,
\ee
where $0^{\pm}$ represents the limit of $x\to0$ from either the right or the left. By symmetry, $u'(0^+) = -u'(0^-)$, so we have on each side of the origin a gradient with magnitude $I$. With this last piece, we arrive at Mornev's result \cref{eqn:I_integral}.

\subsection{Computing the integral in \eqref{eqn:I_integral}}\label{app:I_integral}
For the integral
\be\label{eqn:I_A}
\int_0^{K_D^{n/(n-1)}} du \left(\frac{u^n}{K_D^n + u^n} - u\right),
\ee
since $K_D^{n/(n-1)} \ll K_D$ for $K_D \ll 1$, we can approximate the integrand of the first term by its small $u$ behavior (as $u$ is only being integrated up to a small number) and find
\be
\int du \frac{u^n}{K_D^n + u^n} \approx  \frac{K_D^{-n}u^{n+1}}{n+1}.
\ee

Including the linear term from \eqref{eqn:I_A}, we obtain
\begin{alignat}{2}
\int_0^{K_D^{n/(n-1)}} du \left(\frac{u^n}{K_D^n + u^n} - u\right) &= \frac{1}{n+1}K_D^{\frac{2n}{n-1}} -  \frac{1}{2}K_D^{\frac{2n}{n-1}} \nonumber \\
&= \left(\frac{1}{n+1} - \frac{1}{2}\right)K_D^{\frac{2n}{n-1}}.
\end{alignat}
Substituting into \eqref{eqn:I_integral}, we get
\be
I^{wave}_{c} =\left(\frac{n - 1}{n+1}\right)^{1/2}K_D^{\frac{n}{n-1}},
\ee
and converting to dimensional form, we have
\be
I^{wave}_c = \left(1-\frac{2}{n+1}\right)^{1/2}\left(\frac{K_D}{k/\gamma}\right)^{\frac{n}{n-1}}\sqrt{\frac{D}{\gamma}}\,k.
\ee

\section{Numerical scheme}
\label{sec:numerics}
We solved the PDE \cref{eq:pde_appendix} using the Crank-Nicolson method implemented in Python, building on the \texttt{diffrax} package \cite{kidger_neural_2021}. Our simulation code, along with code for reproducing the figures of this paper, is available at \href{https://github.com/dewitt-lab/immunowave}{https://github.com/dewitt-lab/immunowave}. For the ODE model, we used the same solver but had a grid with a single grid point.

To numerically determine the activation thresholds in simulations, we performed binary search.
For a given stimulus strength, we integrate the model to a steady state, which is considered activated (unactivated) if mean $u$ over the spatial domain is greater (less) than $0.5$.
We initialize with lower and upper bounds for the critical stimulus, first checking their steady states, and extended the bounds, if necessary.
Then we evaluate activation using stimulus strength at the midpoint of the bounding interval.
If this midpoint results in an activated (unactivated) steady state, it becomes the new upper (lower) bound.
This step is iterated until the size of the bounding interval is below a tolerance, and the midpoint is taken as the estimated activation threshold. Simulations were run on a custom-built computer with an Intel Core i9 11900K processor and 128 GB RAM running Ubuntu 20.04. 

\bibliography{references}
\bibliographystyle{unsrtnat} 

\end{document}